\documentstyle[12pt]{article}

\newcommand{\ABO}{\mbox{A $+$ B $\to$ 0}}
\newcommand{\RA}{\rho_{A}}
\newcommand{\RB}{\rho_{B}}
\newcommand{\TRA}{\tilde\rho_{A}}
\newcommand{\TRB}{\tilde\rho_{B}}
\newcommand{\TR}{\tilde{R}}
\newcommand{\PX}[1]{
   \frac{\textstyle\partial{#1}}{\textstyle\partial x}}
\newcommand{\PXX}[1]{
   \frac{\textstyle\partial^2{#1}}{\textstyle\partial x^2}}
\newcommand{\PT}[1]{
   \frac{\textstyle\partial{#1}}{\textstyle\partial t}}

\newcommand{\BRA}[1]{\!\left( #1 \right)}
\newcommand{\SQR}[1]{\!\left[ #1 \right]}
\newcommand{\erf}[1]{ \:\!\mbox{erf}\left(#1\right)}

\newcommand{\ierfs}[1]{\:\!\mbox{erf$^{-1}$}\left[#1\right]}
\newcommand{\GR}{G\'alfi and R\'acz}

\newcommand{\ii}{{\sf ii)}}
\newcommand{\iii}{{\sf iii)}}
\newcommand{\iv}{{\sf iv)}}

\begin{document}

% ***********************
% *                     *
% *     TITLEPAGE       *
% *                     *
% ***********************

\thispagestyle{empty}
\begin{titlepage}
 \begin{center}
   {\LARGE
     The long time behavior of initially separated \ABO\
         reaction-diffusion systems \\
         with arbitrary diffusion constants
     \par
   }
   \vskip 3em
   {\large
     \lineskip .75em
         Zbigniew Koza \\
   }
   \vskip 2em

   {
     \em Institute of Theoretical Physics, University of Wroc\l{}aw, \\
      pl.\ Maxa Borna 9, PL-50204 Wroc\l{}aw, Poland.\\
      zkoza@ift.uni.wroc.pl
   }

\vspace{1em}
  15 November 1995\\
  revised: 18 January 1996\\
  to appear in J. Stat. Phys.

\end{center}

\leftline{\bf Abstract}

\noindent
   We examine the long time behaviour of \ABO\ reaction diffusion
   systems with initially segregated species A and B. All of our
   analysis is carried out for arbitrary (positive) values of the
   diffusion constants $D_A$, $D_B$, and initial concentrations $a_0$
   and $b_0$ of A's and B's. We divide the domain of the partial
   differential equations describing the problem into several regions
   in which they can be reduced to simpler, solvable equations, and we
   merge the solutions. Thus we derive general formulae for the
   concentration profiles outside the reaction zone, the location of
   the reaction zone center, and the total reaction rate. An
   asymptotic condition for the reaction front to be stationary is
   also derived. The properties of the reaction layer are studied in
   the mean-field approximation, and we show that not only the scaling
   exponents, but also the scaling functions are independent of $D_A$,
   $D_B$, $a_0$ and $b_0$.

\vspace{1em}

\vspace{1em}

\end{titlepage}

% ***********************
% *                     *
% *   INTRODUCTION      *
% *                     *
% ***********************

\setcounter{page}{1}
\section{Introduction}
The study of the interfacial region formed in diffusion limited \ABO\
type reactions between domains of unlike species has attracted much
current interest \cite{Overview}-\cite{KozaHaim}. A natural way to
examine this problem is to prepare a system with the components
initially segregated along the plane $x=0$, and then investigate the
spatio-temporal evolution of their concentrations $\RA$ and $\RB$, and
the reaction rate $R$. Such geometry, first studied by \GR\
\cite{G-R}, was already investigated by means of various methods,
including experiments \cite{Experiment,HaimExperiment,HaimKudowa},
numerical simulations \cite{J-E,C95,CDC92,CD-mn}, analytical
computations \cite{Haim91,Bstatic,Linear}, scaling
\cite{G-R,CD-mn,CD-Steady,Cornell} and dimensional
\cite{CD-Steady,hep} analysis.

A standard way to treat the initially separated problem analytically
is to solve the following partial differential equations \cite{G-R}

\begin{equation}
   \label{GR}
    \left.
     \begin{array}{rcl}
       \PT{\RA} &=& D_A \PXX{\RA} - R\\[2ex]
       \PT{\RB} &=& D_B \PXX{\RB} - R
     \end{array}
     \right\}
\end{equation}
with the initial state given by
\begin{equation}
  \label{IniCond}
   \left.
    \begin{array}{l}
      \RA(x,t=0) \;=\; a_0 H(-x) \\[1ex]
      \RB(x,t=0) \;=\; b_0 H(x)
    \end{array}
  \right\}
\end{equation}
where $\RA(x,t)$ and $\RB(x,t)$ are the local concentrations of A's
and B's, $R$ is the reaction rate, $H(x)$ denotes the Heavyside step
function, and $a_0$, $b_0$, $D_A$ and $D_B$ are some positive
constants related to the initial concentrations of species A and B and
their diffusion coefficients respectively. It is customary
\cite{G-R,CDC92,CD-mn,Linear,CD-Steady,Cornell,hep,BenRedner,Kudowa-H}
to assume $D_A = D_B \equiv D$, which leads to the conclusion that
$u(x,t) \equiv \RA - \RB$ obeys the readily solvable diffusion
equation $\partial_t u = D \partial^{2}_{x}u$ irrespective of $R$.
Finally some form of $R$ must be assumed, and in most cases either the
mean field approximation $R \propto \RA\RB$
\cite{G-R,Haim91,Linear,BenRedner}, or its generalization $R \propto
\RA^{m}\RB^{n}$ \cite{CDC92,CD-mn,CD-Steady,Cornell} was adopted.

With these assumptions, two fundamental concepts were developed, both
referring to the long time limit. According to the first one
\cite{G-R}, the long time behavior of the system inside the reaction
layer can be described with a help of some scaling functions $S_A$,
$S_B$ and $S_R$ through

\begin{eqnarray}
  \label{4}
    \RA(x,t) &\propto&
         t^{-\gamma}S_A\BRA{x - x_f(t) \over t^\alpha}\;,\\[1ex]
  \label{5}
    \RB(x,t) &\propto&
         t^{-\gamma}S_B\BRA{x - x_f(t) \over t^\alpha}\;,\\[1ex]
  \label{6}
      R(x,t) &\propto&
         t^{-\beta}  S_R\BRA{x - x_f(t) \over t^\alpha}\;,
\end{eqnarray}
where $x_f(t)$ denotes the point at which the reaction rate $R$
attains its maximal value, and exponents $\alpha$, $\beta$ and
$\gamma$ are some positive constants given, for $R \propto
\RA^{m}\RB^{n}$, by $\gamma = 1/(m+n+1)$, $\alpha = \frac{1}{2} -
\gamma$ and $\beta = 1-\gamma$ \cite{CD-mn}. The scaling ansatz is
based on the assumption that the width $w(t)$ of the reaction layer
grows with time as $t^\alpha$ with $\alpha < 1/2$, so that in addition
to the diffusion length scale $\lambda_D \sim \sqrt{Dt}$, the problem
possesses also another relevant length scale $w \propto t^\alpha$.

According to the second theory, called the quasistationary
approximation \cite{CD-Steady,BenRedner}, the currents $J_{A}(t)$ and
$J_{B}(t)$ of particles A and B arriving at the interface layer from
the two densely occupied domains are changing so slowly, that the
relatively narrow interface has enough time to equilibrate. To
'equilibrate' means here to reach a state completely determined by the
current boundary conditions, i.\ e.\ by $J_A$ and $J_B$.
Mathematically this is equivalent to the assumption that the state of
the reaction zone is entirely given by equations obtained from
(\ref{GR}) by replacing their left sides, or the time derivatives,
with zero. This leads to much simpler equations

\begin{equation}
   \label{STAC}
    \left.
     \begin{array}{rcl}
       D_A \PXX{\RA} &=& R\\[2ex]
       D_B \PXX{\RB} &=& R
     \end{array}
     \right\}
\end{equation}
which are to be solved with the boundary conditions
$\partial\RA/\partial x \to -J_A(t)$ and $ \RB \to 0$ as $x \to
-\infty$, and $\RA \to 0$, $\partial\RB/\partial x \to J_B(t)$ as $x
\to +\infty$. The most important feature of the quasistationary
equations (\ref{STAC}) is that they depend only on $x$, with time $t$
being a parameter entering their solutions $\RA(x,t)$ and $\RB(x,t)$
only through the time dependent boundary currents $J_A$ and $J_B$.

It was conjectured by \GR\ \cite{G-R} that the first of the above
assumptions, $D_A = D_B$, is irrelevant with regard to the long time
behavior of the system, the ratio $D_A/D_B$ affecting perhaps the form
of the scaling functions $S_A$, $S_B$ and $S_R$, but not the values of
exponents $\alpha$, $\beta$ and $\gamma$. This hypothesis was
generally accepted after numerical \cite{J-E} and experimental
\cite{Experiment} verification. However, there is still no analytical
theory referring to the general case $D_A \neq D_B$. For two reasons
this situation arouses some anxiety. First, it is practically
impossible to find in Nature two species with exactly the same
diffusion constants. Second, the above mentioned verification
encompassed only the case where the ratio $D_A/D_B$ was of order 1,
whereas it is known \cite{J-E,Kudowa-H}, that if one of the diffusion
constants is equal zero, the mean-field exponents assume values
entirely different from those predicted by \GR, namely $\alpha = 0$,
$\beta = 1/2$ and $\gamma = 1/4$. The aim of this paper is to present
such general theory comprising the case of any positive diffusion
constants $D_A$ and $D_B$.

Unfortunately, we know of only one successful attempt to derive the
macroscopic form of $R$ from the microscopic properties of the system
\cite{Bstatic}. Dimensional analysis leads to another important
conclusion that the mean field approximation should be valid only in
spaces of dimension higher than $d_c = 2$ \cite{CD-mn,CD-Steady,hep}.
Therefore our basic equation (\ref{GR}) might seem useful only for
these two sorts of systems for which the form of $R$ is known. In our
approach, however, we will not need to impose any special restriction
on the form of $R$. Instead, we will require that the solutions of
(\ref{GR}) satisfy a few physically justifiable relations. Therefore
our theory can be applied even to the systems for which the form of
$R$ remains unknown, including experiments and microscopic models. In
such cases verification of our postulates should be far easier
than the task of finding the exact form of $R$, let alone solving
(\ref{GR}) afterwards.

The paper is organized as follows. In the next section we will present
the assumptions our theory has been founded on, as well as their short
physical justification. The general theory is formulated in the third
section. In the next section we will use it to derive and discuss the
scaling ansatz in the mean field approximation. The final, fifth
section is devoted to conclusions.

% ***********************
% *                     *
% *    ASSUMPTIONS      *
% *                     *
% ***********************

\section{Assumptions}

We will consider systems which can be described with the \GR\
equations (\ref{GR}) and the boundary conditions (\ref{IniCond}). We
will assume that $D_A$, $D_B$, $a_0$ and $b_0$ are some known positive
constants. Our analysis will be based on a few physical assumptions.

\renewcommand{\theenumi}{\roman{enumi}}
\begin{enumerate}
 \item At any time $t > 0 $ there exists a unique point $x_f(t)$ at
       which the reaction term $R$ attains its maximal value, and a
       unique point $x_0(t)$ at which $D_A\RA(x_0,t) - D_B\RB(x_0,t) =
       0$.

 \item The reaction is concentrated in a region $|x - x_f| \sim w(t)
      \sim t^{\alpha}$ with $0 < \alpha < 1/2$.  Outside this region,
      for $x \ll x_f - w$, there is $\RA \gg \RB$, and for $x \gg x_f
      + w$ we have $\RA \ll \RB$.

 \item The evolution of $\rho_A$ in the region $x \ll x_f - w$
  can be approximated by
  \begin{equation}
         \label{ra}
   \RA(x,t) \; = \; a_0 -  C_A\SQR{ \erf{x / \sqrt{4D_At}} + 1}\;,
  \end{equation}
  where $C_A$ is a constant, and \hspace{0.3ex} $\erf{x} \equiv
  2\pi^{-1/2}\int_{0}^{x} \exp(-\eta^2)\!\;\mbox{d}\eta$
  \hspace{0.3ex} is the error function \cite{Luke}.

  Similarly, for $x \gg x_f + w$, the evolution of $\RB$ can
  be estimated by
   \begin{equation}
         \label{rb}
     \RB(x,t) \;=\;  b_0 + C_B \SQR{\erf{ x / \sqrt{4D_B t}} - 1}\;,
   \end{equation}
  where $C_B$ denotes another constant. Both $C_A$ and $C_B$ depend on
  the initial parameters $a_0$, $b_0$, $D_A$ and $D_B$.

  \item The quasistatic approximation is valid in the region
        $-(D_A t)^{1/2} \ll x \ll (D_B t)^{1/2}$.

\end{enumerate}

The first assumption introduces two functions $x_f(t)$ and $x_0(t)$,
restricting the considerations to the cases where they are uniquely
defined. Function $x_f$ identifies directly the location of the
reaction layer at time $t$, and $x_0$ is an auxiliary, mathematical
object helpful in examining the behavior of $x_f$. That $x_0$ exists
for any $t>0$  stems from the initial conditions (\ref{IniCond}). As
for the second postulate, it was satisfied by all the \ABO\
interfacial systems examined so far. The third assumption comes from
the observation that, due to postulate \ii, in the region $x \ll x_f -
w$ the concentration of particles A is expected to be much bigger than
that of B's, the latter having to cross the whole reaction layer to
get there. Therefore, the evolution of A's is  practically unaffected
by B's, and so it should be governed by the standard diffusion
equation $\partial_t\RA = D_A\partial^2_x{\RA}$. The particular, based
on the error function form (\ref{ra}) of its solution  was predicted
and experimentally confirmed by Koo and Kopelman \cite{Experiment}.
Notice also that for any time $t$ such form of $\RA$ guarantees that
the relation $\lim_{x\to-\infty}\RA = a_0$ implied by the initial
conditions (\ref{IniCond}) is also fulfilled. A similar argument leads
to (\ref{rb}). As for the last postulate, the quasistationary
approximation is based on the following observation \cite{CD-Steady}.
The diffusion current of particles arriving at the reaction layer is
$J \propto t^{1/2}$, so the characteristic time scale on which this
current changes is $\tau_J \sim (\mbox{d}\log J/ \mbox{d} t)^{-1}
\propto t$, whereas the equilibration time of the reaction front is
$\tau_F \sim w^2 \propto t^{2\alpha}$; therefore $\alpha < 1/2$
implies that as time goes to infinity, the ratio $\tau_F/\tau_J$ goes
to $0$, validating the quasistatic approximation.

As we mentioned above, we will not impose any explicit restrictions on
the form of the macroscopic reaction rate $R$ requiring only that it
be consistent with the above postulates.
However, to investigate the behavior of the \ABO\ system inside
the reaction zone we will need more detailed information about $R$.
Therefore in Section 4 we will concentrate on the mean-field
approximation $R \propto \RA\RB$.

% ***********************
% *                     *
% *     ANALYSIS        *
% *                     *
% ***********************

\section{Analysis}

\begin{figure}[htpb]
\thicklines
 \begin{center}
  \unitlength1mm
  \begin{picture}(120,32)(0,12)
  \put(9,20){\line(1,0){28}}
  \put(40,20){\makebox(0,0){\ldots}}
  \put(43,20){\line(1,0){39}}
  \put(85,20){\makebox(0,0){\ldots}}
  \put(88,20){\line(1,0){24}}
  \put(4,20){\makebox{\ldots}}
  \put(4,17){\makebox(0,0){$-\infty$}}
  \put(118,20){\makebox(0,0)[r]{\ldots}}
  \put(118,17){\makebox(0,0)[r]{$\infty$}}
  \put(60,18){\line(0,1){2}}
  \put(60,17){\makebox(0,0)[t]{$x_f$}}
  \put(50,18){\line(0,1){2}}
  \put(50,17){\makebox(0,0)[t]{$x_f-w$}}
  \put(70,18){\line(0,1){2}}
  \put(70,17){\makebox(0,0)[t]{$x_f+w$}}
  \put(30,18){\line(0,1){2}}
  \put(30,17){\makebox(0,0)[t]{$-\sqrt{D_A t}$}}
  \put(100,18){\line(0,1){2}}
  \put(100,17){\makebox(0,0)[t]{$\sqrt{D_B t}$}}
  \put(27,24){\makebox(0,0)[b]{\shortstack{
        \small \bf iii \\
        \small free diffusion of A's
         }}}
  \put(27,22){\vector(-1,0){23}}
  \put(27,22){\vector(1,0){23}}
  \put(94,24){\makebox(0,0)[b]{\shortstack{
        \small \bf iii \\
        \small free diffusion of B's
        }}}
  \put(94,22){\vector(-1,0){24}}
  \put(94,22){\vector(1,0){24}}

  \put(65,35){\makebox(0,0)[b]{\shortstack{
       \small \bf iv \\
       \small quasistationary approximation
        }}}
  \put(65,33){\vector(-1,0){35}}
  \put(65,33){\vector(1,0){35}}
  \end{picture}
 \end{center}
 \caption{\sf Schematic diagram of the regions of applicability of
  postulates \iii\ and \iv.
  Asymptotically  $w(t)  \propto t^{\alpha} \ll t^{1/2}$.
 }
 \label{fig1}
\end{figure}
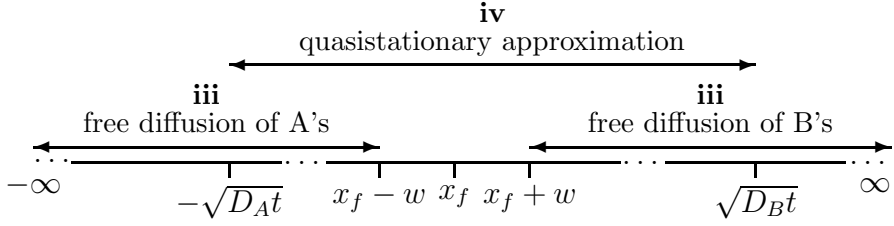

The following observation constitutes the basis of the analysis of our
model. For sufficiently long time $t$, at any point $x$ we can employ
either assumption \iii\ or \iv\ or  both of them -- see Fig.\
\ref{fig1}. Therefore we can divide the $x$ axis into several regions,
and in each of them the initial problem of solving (\ref{GR}) can be
reduced to a much simpler one. Then, the overlapping of the domains of
applicability of \iii\ and \iv\ will enable us to merge the solutions.

% ***********************
% *                     *
% *       FIG. 1        *
% *                     *
% ***********************

%%%%%%%%%     FIG 1     %%%%%%%%%%%

%\typeout{Processing Fig. 1}

% ***************
% *             *
% *  END FIG1   *
% *             *
% ***************

Consider first the region $-\sqrt{D_A t} \ll x \ll \sqrt{D_B t}$. By
assumption \iv\ the system is governed here by quasistationary
equations (\ref{STAC}). They imply that $\Psi (x,t) \equiv D_B\RB -
D_A\RA $ satisfies $\partial^2\Psi/\partial x^2 = 0$. Therefore $\Psi$
is linear in $x$. Let $J(t)$ denote its slope. By definition of $x_0$
we have $\Psi(x_0,t) = 0$. Thus we arrive at the conclusion that at
sufficiently long time $t$, for $-\sqrt{D_A t} \ll x \ll \sqrt{D_B
t}$, there is
\begin{equation}
 \label{defj}
  D_B\RB - D_A\RA \;\approx\; J(t) (x-x_0(t)) \;,
\end{equation}
and so $J_A(t) = J_B(t) = J(t)$.  The notation
$f(t) \approx g(t)$ means $\lim_{t\to\infty} f(t)/g(t) = 1$.

Consider now the region $-\sqrt{D_A t} \ll x \ll x_f - w$, so that
$\epsilon \equiv x_f - x$ fulfils $t^{\alpha} \ll \epsilon \ll
t^{1/2}$.
Applying assumption \ii\ to (\ref{defj}) we can approximate the
form of $\RA$ by
\begin{equation}
   \label{raj}
  \RA(x,t) \;\approx\; -D_A^{-1}J(t) (x-x_0(t)) \;.
\end{equation}
On the other hand, however, by assumption \iii, $\RA$ can be here as
well expressed by equation (\ref{ra}). So we have
\begin{equation}
     \label{eps1}
    a_0 - C_A \SQR{\erf{\frac{x_f(t)-\epsilon}{\sqrt{4D_At}}} + 1}
          \;\approx\; -D_A^{-1}J(t)(x_f(t) - x_0(t) - \epsilon)
\end{equation}
and
\begin{equation}
  \label{dereps}
   \left.
   \PX{}\BRA{a_0 - C_A\SQR{\erf{\frac{x}{\sqrt{4D_At}}}+1}}
       \right|_{x_f - \epsilon}
       \;\approx\;
   -D_A^{-1}J(t) \;.
\end{equation}

By assumption \ii, for any $x$ located outside the reaction layer, the
ratio $\RA/\RB$ will either converge to zero, or diverge to infinity
as $t\to\infty$. However, by definition of $x_0$, this ratio assumes
the constant value $D_B/D_A$ at $x =x_0$. So $x_0$ must lie inside the
reaction layer. As its width grows as $t^{\alpha}$, we conclude that
there must exist a number $\theta$ such that $|x_f(t) - x_0(t)| \le
\theta t^{\alpha}$. We can see now that in the long time limit $|x_f -
x_0|$ becomes negligibly small compared to $\epsilon$ which, in turn,
gets negligibly small compared to $t^{1/2}$. Therefore we can drop
$\epsilon$ on the left hand side of (\ref{eps1}) and (\ref{dereps}),
and $x_f - x_0$ on the r.\ h.\ s.\ of (\ref{eps1}). After these
transformations the asymptotic value of the l.\ h.\ s.\ of
(\ref{eps1}) turns out independent of $\epsilon$, whereas the r.\ h.\
s.\ of (\ref{eps1}) becomes proportional to $\epsilon J(t)$. As
$\epsilon$ can vary between $t^\alpha$ and $t^{1/2}$, we conclude that
$J(t)\epsilon(t)$ goes either to 0, or to $\infty$. The latter case is
impossible because (\ref{eps1}) approximates the value of $\RA$ which
must be finite. In the long time limit we therefore have
\begin{equation}
  \label{epsj0}
   J(t)  \epsilon(t) \;\to\;  0 \;,
\end{equation}
\begin{equation}
      \label{aat0}
  a_0 - C_A \SQR{\erf{\frac{x_f(t)}{\sqrt{4D_At}}} + 1} \;\to\; 0 \;,
\end{equation}
and
\begin{equation}
  \label{der2a}
  J(t)\sqrt{t} \;\to\;
  C_A\sqrt{ D_A/\pi} \exp\BRA{-\frac{x_f^2(t)}{4D_At}} \;.
\end{equation}

Similar arguments applied to the region $x_f + w \ll x \ll \sqrt{D_B
t}$ lead to
\begin{equation}
      \label{bat0}
  b_0 + C_B \SQR{\erf{\frac{x_f(t)}{\sqrt{4D_B t}}} - 1} \;\to\; 0 \;,
\end{equation}
and
\begin{equation}
  \label{der2b}
  J(t)\sqrt{t} \;\to\; C_B\sqrt{D_B/\pi}
             \exp\BRA{-\frac{x_f^2(t)}{4D_B t}} \;.
\end{equation}

It follows from (\ref{aat0}) and (\ref{bat0}) that in the
long time limit
\begin{equation}
  \label{defC_f}
  x_f(t)/\sqrt{t} \;\to\;  C_f \;,
\end{equation}
where $C_f$ is a constant given either by
\begin{equation}
  \label{Cf1}
  C_f \;=\; 2\sqrt{D_A}\ierfs{(a_0 - C_A)/C_A}
\end{equation}
or
\begin{equation}
  \label{Cf2}
  C_f \;=\; 2\sqrt{D_B}\ierfs{(C_B-b_0)/C_B} \;.
\end{equation}

Now  (\ref{der2a}), (\ref{der2b}) and (\ref{defC_f}) imply that as
time goes to infinity we have
\begin{equation}
  \label{defC_J}
  J(t)\sqrt{t} \;\to\;  C_J \;,
\end{equation}
where $C_J$ is another constant given either by
\begin{equation}
  \label{CJ1}
  C_J \;=\; C_A\sqrt{D_A/\pi}\exp\BRA{-\frac{C_f^2}{4D_A}}
\end{equation}
or
\begin{equation}
  \label{CJ2}
  C_J \;=\; C_B\sqrt{D_B/\pi}\exp\BRA{-\frac{C_f^2}{4D_B}} \;.
\end{equation}
Notice that (\ref{defC_J}) is consistent with (\ref{epsj0}).

So far we have introduced four constants $C_A$, $C_B$,  $C_f$ and
$C_J$. The first two of them, $C_A$ and $C_B$, control the asymptotic
profile of the majority species outside the reaction layer. The third
constant, $C_f$, governs the location of the reaction layer center.
Finally, through the formula $J(t) \approx \int\! R(x,t) \mbox{d}x
\approx C_J/t^{1/2}$, parameter $C_J$ is related to the magnitude of
the current $J(t)$ of particles entering the reaction layer, or,
equivalently, the total reaction rate at time $t$. Due to the form of
the initial state (\ref{IniCond}) we expect $\partial_x \RA \le 0$ and
$\partial_x \RB \ge 0$, which implies $C_A > 0$, $C_B > 0$ and $C_J >
0$.

Equations (\ref{Cf1}), (\ref{Cf2}), (\ref{CJ1}) and (\ref{CJ2})
can be reduced to
\begin{equation}
  \label{1eq}
   \Phi\!\left(
             \frac{-C_f}{2\sqrt{D_A}}
       \right)
   \;=\;
        \frac{a_0\sqrt{D_A}}{b_0\sqrt{D_B}} \:
        \Phi\!\left(
                  \frac{C_f}{2\sqrt{D_B}}
            \right) \;,
\end{equation}
where
\begin{equation}
  \label{defphi}
  \Phi (x) \;\equiv\; \SQR{1 - \erf{x}}\exp(x^{2}) \;.
\end{equation}
An important feature of $\Phi(x)$ is that it diminishes monotonically
from $\infty$ to $0$ as $x$ grows from $-\infty$ to $\infty$. This
property guarantees that equation (\ref{1eq}) always has a unique
solution $C_f = C_f(a_0/b_0, D_A, D_B)$ which, moreover, can be
readily found numerically. The only problem that can appear while
solving (\ref{1eq}) numerically is that when $x$ is positive,
$\Phi(x)$ is a product of a very small and a very big numbers. For
this reason, if $x$ is greater than 5, we suggest to use the
asymptotic form $\Phi(x) \approx 1/(\sqrt{\pi}x)$ which comes from the
asymptotic properties of the error function \mbox{erf} \cite{Luke}.

With $C_f$ computed from (\ref{1eq}), the values of  $C_A$, $C_B$ and
$C_J$ can now be calculated from (\ref{Cf1}), (\ref{Cf2}) and
(\ref{CJ1}). The opposite statement is also true: if we know (e.\ g.\
from an experiment) the values of $C_A$, $C_B$, $C_f$ and $C_J$, our
equations determine uniquely the values of $a_0$, $b_0$, $D_A$ and
$D_B$.

The immediate consequence of (\ref{1eq}) is that the sign of $C_f$ is
determined by the sign of $a_0\sqrt{D_A}/(b_0\sqrt{D_B}) - 1$. In
particular we conclude that

\begin{equation}
  \label{Cf=0}
  C_f \;=\; 0 \;\; \iff \;\;
  a_0\sqrt{D_A} \;=\; b_0\sqrt{D_B} \;.
\end{equation}
This formula is important for planning experiments, as it clarifies
the way the initial concentrations of the species should be chosen in
order to have the reaction layer move asymptotically as slowly as
possible. Condition  (\ref{Cf=0}) is consistent with that of Jiang and
Ebner's \cite{J-E} who, by numerical examination of the mean field
approximation $R\propto \RA\RB$, found a stronger relation $x_f \,=\,
0 \;\iff\; a_0\sqrt{D_A} \,=\, b_0\sqrt{D_B}$. Our  general formula,
derived for any reaction term $R$, implies only that with this
particular choice of the initial parameters the function $x_f$ cannot
be changing as fast as $t^{1/2}$. An example of a system where $C_f =
0$ and $x_f(t) \propto t^{\alpha}$ was investigated in \cite{Cornell}.

Equation (\ref{1eq}) enables us also to observe a striking similarity
between the long and short time behavior of $x_f$. According to
\cite{Haim91}, in the short time limit the reaction term does not
affect the solutions of (\ref{GR}), and so $\RA$ and $\RB$ assume the
same forms as in the readily solvable case $R = 0$. The point $x_f$
can be then found as the point at which $\partial R/\partial x = 0$.
For $R \propto \rho_A^m \rho_B^n$ such procedure yields $\lim_{t\to 0}
\, x_f/\sqrt{t} = C_0$, where $C_0$ can be found from the relation
very similar to that of (\ref{1eq})

\begin{equation}
  \label{phi2}
   \Phi\!\left(
             \frac{C_0}{2\sqrt{D_A}}
       \right)
   \;=\;
        \frac{m\sqrt{D_B}}{n\sqrt{D_A}} \:
        \Phi\!\left(
                  \frac{-C_0}{2\sqrt{D_B}}
            \right) \;.
\end{equation}

% ***********************
% *                     *
% *   REACTION LAYER    *
% *                     *
% ***********************

\section{The reaction layer}

In the previous section we carried out our analysis without imposing
any restrictions on the form  of the macroscopic reaction term $R$. As
we now proceed to examine the asymptotic properties of the reaction
layer, we will obviously need more specific information about $R$.
Therefore we will concentrate on the mean field approximation $R =
k\RA\RB$, $k= \mbox{const.}$, still allowing $a_0$, $b_0$,
$D_A$ and $D_B$ to take any positive values.

By assumption \iv\ we expect that in the region $-(D_A t)^{1/2} \ll x
\ll (D_B t)^{1/2}$ we can apply the quasistatic approximation
equations (\ref{STAC}). Let $\RA(x,t)$ and $\RB(x,t)$ denote their
solutions for some values of $D_A$, $D_B$, $x_0(t)$ and $J(t)$. By the
following linear transformation we introduce two new functions of a
single variable $\TRA(x)$ and $\TRB(x)$
\begin{equation}
  \label{Solutions}
  \left.
    \begin{array}{rcl}
      \RA(x,t) &=& \eta_A (t)\TRA\SQR{(x-x_0(t))/w(t)}\\[1.5ex]
      \RB(x,t) &=& \eta_B (t)\TRB\SQR{(x-x_0(t))/w(t)}
    \end{array}
  \right\}
\end{equation}
where
\begin{eqnarray}
  \label{defw}
  w(t) &\equiv&     \sqrt[3]{\frac{D_A D_B}{kJ(t)}}
       \;=\;    \sqrt[3]{\frac{D_A D_B}{kC_J}}\,t^{1/6} \;, \\[1ex]
  \label{etaa}
  \eta_A(t) &\equiv& J(t)w(t)/D_A
       \;=\;
       \BRA{\frac{D_B}{k}}^{1/3}\BRA{\frac{C_J}{D_A}}^{2/3}t^{-1/3}
         \;,\\[1ex]
  \label{etab}
  \eta_B(t) &\equiv& J(t)w(t)/D_B
       \;=\;
       \BRA{\frac{D_A}{k}}^{1/3}\BRA{\frac{C_J}{D_B}}^{2/3}t^{-1/3}
         \;.
\end{eqnarray}
Denoting $\TR(x) \equiv \TRA(x)\TRB(x)$ we have also
\begin{equation}
  \label{R}
  R(x,t) \;\equiv\; k\RA\RB \;=\; C_J^{4/3}(D_A D_B)^{-1/3} k^{1/3}
  t^{-2/3} \TR\SQR{(x-x_0)/w(t)} \;.
\end{equation}

The essential property of $\TRA(x)$ and $\TRB(x)$ is that they
constitute the particular solution to equations (\ref{STAC}) with $D_A
= D_B = J = k = 1$ and $x_0 = 0$. Therefore, by symmetry, $\TR(x)$
assumes its maximal value for $x = 0$, and so equation (\ref{R})
implies that $R$ attains the maximal value at $x = x_0$. In the long
time limit we can therefore identify $x_f$ with $x_0$. Comparing now
(\ref{Solutions}) and (\ref{R}) with the scaling ansatz (\ref{4}) -
(\ref{6}) we see that we can also identify $\TRA$ with $S_A$, $\TRB$
with $S_B$, and $\TR$ with $S_R$. Therefore not only are the above
formulae consistent with \GR's scaling ansatz, but through $C_J(a_0,
b_0, D_A, D_B)$ they also {\em exactly} relate the quantities of
physical importance (e.\ g.\ $w(t)$) to the parameters of the system
($D_A$, $D_B$, $a_0$, $b_0$ and $k$).

Because (\ref{Solutions}) can be applied to systems with any positive
values of 'external' parameters $a_0$, $b_0$, $D_A$, $D_B$ and $k$, we
arrive at the conclusion that the long time evolution of initially
segregated \ABO\ systems is even more universal that it was predicted
by \GR; namely, not only the scaling exponents, but also the form of
the scaling functions does not depend on the external parameters.
Therefore, to find the scaling properties of the reaction layer it is
sufficient to concentrate on the simplest, symmetric case $D_A = D_B$
and $a_0 = b_0$.

Notice that we have achieved these results by means of a simple,
linear transformation (\ref{Solutions}). In this way we took advantage
of the very feature of equations (\ref{GR}) and (\ref{STAC}) that
prevents them from being solved analytically -- nonlinearity.

The above analysis is straightforward and can be easily generalized
for many other reaction terms $R$. In particular, for $R =
k\RA^m\RB^n$, with $k = \mbox{const}$ and $m$, $n$ being any
(positive) real numbers, the following relation should be used instead
of (\ref{defw})

\begin{equation}
  \label{defwQ}
  w^{m+n+1} \;\equiv\; D_A^m D_B^n k^{-1} J^{1-m-n}
            \;.
\end{equation}
This formula, together with (\ref{defC_J}), (\ref{etaa}) and
(\ref{etab}), generalizes the scaling theory of Cornell {\em et al}
\cite{CD-mn} for the case of any positive $a_0$, $b_0$, $D_A$, $D_B$,
$m$ and $n$.

% ***********************
% *                     *
% *     CONCLUSIONS     *
% *                     *
% ***********************

\section{Conclusions}

We have investigated the long time behavior of the concentrations
$\RA$ and $\RB$ of phases A and B in the \GR's problem. Our analysis
is the first analytical attempt to consider it in the general case of
arbitrary positive initial concentrations $a_0$ and $b_0$, and
diffusion constants $D_A$ and $D_B$ of A's and B's.

Our approach is very general, as it does not impose any restrictions
on the form of the macroscopic reaction rate $R$.  Instead, it is
based on the assumption that in the long time limit $\RA$ and $\RB$
satisfy a few physically justifiable relations. Therefore our theory
can be applied to various systems, including those for which the form
of the macroscopic reaction rate $R$ remains unknown. Another peculiar
feature of
our theory is that, unlike most of previous studies,
it does not concentrate on the investigation of the reaction layer
only, but takes into account the properties of the whole, infinite
system.

In this way we managed to derive general formulae for the
concentration profiles of the majority species outside the reaction
layer, the location of the layer, and the total reaction rate. It is
interesting to notice that these quantities turned out independent of
$R$. We also derived analytically Jiang and Ebner's condition for the
reaction front to be asymptotically stationary. This relation also
turned out independent of $R$. These results correspond to the recent
findings based on dimensional analysis \cite{CD-mn,CD-Steady,hep},
according to which the scaling properties of the reaction layer are
independent of the form of $R$.

Next we derived the general scaling ansatz for the mean field
approximation. We gave the formulae which exactly relate some
quantities of physical importance, (e.\ g.\ the width $w$ of the
reaction layer) to the external parameters of the system $a_0$, $b_0$,
$D_A$, $D_B$ and $k$. It turned out that not only the scaling
exponents, but also the forms of the scaling functions are independent
of the values of these parameters. This justifies the customary
approach of examining the properties of the reaction layer only in the
simplest, symmetric case $a_0 = b_0$ and $D_A = D_B$.

Our work suggests also that the behavior of the reaction-diffusion
system can be understood as a subtle interplay between two scaling
regimes. The first one is valid far from the reaction zone, where the
densities of particles A and B assume the scaling forms typical of
purely diffusive systems: $\RA(x,t) \approx \Psi_A(x/t^{1/2})$ and
$\RB(x,t) \approx \Psi_B(x/t^{1/2})$. These scaling laws determine
also the location of the point $x_f(t)$ of the maximal reaction, and
the magnitude of the current $J(t)$ of the particles entering the
reaction zone.  However, at $x_f$ the spatial derivatives of $\Psi_A$
and $\Psi_B$ suffer discontinuity. Therefore in the vicinity of $x_f$
a new form of scaling develops, and $\RA$ and $\RB$ assume the form
$\RA(x,t) = S_A(x/t^\alpha)$ and $\RB(x,t) = S_B(x/t^\alpha)$ with
$\alpha < 1/2$.

Although we confined our considerations only to the long time limit,
it would be interesting to combine our results with those of
Taitelbaum {\em et al} \cite{Haim91} for the short and intermediate
times. We believe that the striking similarity between equations
(\ref{1eq}) and (\ref{phi2}) is not accidental and should lead to a
general theory comprising the short, intermediate and long time limit.
The first attempt in this direction has already been made
\cite{KozaHaim}.

Notice also that the quasistationary approximation leads to new
definitions of 'short', 'intermediate' and 'long' time regimes.
Namely, we can define them as the time intervals in which the reaction
term, in the vicinity of $x_f$, is vanishingly small compared to the
time derivative ('short time'); or the interval in which they are of
similar magnitude ('intermediate time'); or the interval in which it
is the time derivative that can be neglected ('long time').

Another interesting problem concerns the limit $D_A \to 0$ with
other external parameters fixed. In this limit the scaling exponents
(in the mean-field approximation) are expected to change from $\alpha
= 1/6$, $\beta = 2/3$ to $\alpha = 0$, $\beta = 1/2$. The paper in
which this problem is examined within the framework of the presented
here theory is under preparation. We will mention here only that as
$D_A $ goes to 0, the time at which the system reaches the long time
regime goes to infinity, so the case $D_A = 0$ can be considered as the
case where the system always remains in the 'intermediate' time regime.

\vspace{3ex}
{\bf Acknowledgments}
I am grateful to prof.\ {\L}.\ A.\ Turski for pointing my attention to
the case of unequal diffusion constants, and dr.\ H.\ Taitelbaum for
many inspiring discussions and hospitality at the Bar-Ilan University.
This work was supported by the Polish KBN Grant nr 2 P 302 181 07.

% ***********************
% *                     *
% *    BIBLIOGRAPHY     *
% *                     *
% ***********************

\end{document}